\documentclass[sigplan,10pt,dvipsnames]{acmart}
\usepackage{amscd,amsmath, cite,multirow}
\usepackage[justification=centering,skip=0pt,belowskip=-15pt,aboveskip=0pt]{caption}
\usepackage{enumitem}
\usepackage{subcaption}
\newtheorem{theorem}{Theorem}

\usepackage{mdframed}

\usepackage{comment}

\newcommand\remove[1]{}
\makeatother

\usepackage{tikz}

\newcommand{\hlcolor}{Yellow!35}
\newcommand{\hlcolorTwo}{LimeGreen!35}
\makeatletter
\newenvironment{btHighlight}[1][]
{\begingroup\tikzset{bt@Highlight@par/.style={#1}}\begin{lrbox}{\@tempboxa}}
{\end{lrbox}\bt@HL@box[bt@Highlight@par]{\@tempboxa}\endgroup}

\newcommand\btHL[1][]{%
  \begin{btHighlight}[#1]\bgroup\aftergroup\bt@HL@endenv%
}
\def\bt@HL@endenv{%
  \end{btHighlight}%
  \egroup
}
\newcommand{\bt@HL@box}[2][]{%
  \tikz[#1]{%
    \pgfpathrectangle{\pgfpoint{1pt}{0pt}}{\pgfpoint{\wd #2}{\ht #2}}%
    \pgfusepath{use as bounding box}%
    \node[anchor=base west, fill=\hlcolor,outer sep=0pt,inner xsep=1pt, inner ysep=0pt, rounded corners=2pt, minimum height=\ht\strutbox+2pt,#1]{\raisebox{1pt}{\strut}\strut\usebox{#2}};
  }%
}

\newenvironment{btHighlightTwo}[1][]
{\begingroup\tikzset{bt@HighlightTwo@par/.style={#1}}\begin{lrbox}{\@tempboxa}}
{\end{lrbox}\bt@HLTwo@box[bt@HighlightTwo@par]{\@tempboxa}\endgroup}

\newcommand\btHLTwo[1][]{%
  \begin{btHighlightTwo}[#1]\bgroup\aftergroup\bt@HLTwo@endenv%
}
\def\bt@HLTwo@endenv{%
  \end{btHighlightTwo}%
  \egroup
}
\newcommand{\bt@HLTwo@box}[2][]{%
  \tikz[#1]{%
    \pgfpathrectangle{\pgfpoint{1pt}{0pt}}{\pgfpoint{\wd #2}{\ht #2}}%
    \pgfusepath{use as bounding box}%
    \node[anchor=base west, fill=\hlcolorTwo,outer sep=0pt,inner xsep=1pt, inner ysep=0pt, rounded corners=2pt, minimum height=\ht\strutbox+2pt,#1]{\raisebox{1pt}{\strut}\strut\usebox{#2}};
  }%
}

\usepackage{listings}
\usepackage{etoolbox}
\newtoggle{InString}{}
\togglefalse{InString}
\newcommand*{\ColorIfNotInString}[1]{\iftoggle{InString}{#1}{\color{blue}#1}}%
\newcommand*{\ProcessQuote}[1]{#1\iftoggle{InString}{\global\togglefalse{InString}}{\global\toggletrue{InString}}}%
\definecolor{code_indent}{HTML}{CCCCCC}

\usepackage{multicol}

\acmConference[PL'18]{ACM SIGPLAN Conference on Programming Languages}{January 01--03, 2018}{New York, NY, USA}
\acmYear{2018}
\acmISBN{} 
\acmDOI{} 
\startPage{1}
\setcopyright{none}
\begin{document}

\title{Parallel Batched Interpolation Search Tree}
\author{Vitaly Aksenov}
\affiliation{ITMO University}

\author{Ilya Kokorin}
\affiliation{ITMO University}

\author{Alena Martsenyuk}
\affiliation{MIPT}
\date{August 2021}

\begin{abstract}
Ordered set (and map) is one of the most used data type. In addition to standard set operations, like \texttt{insert}, \texttt{delete} and \texttt{contains}, it can provide set-set operations such as \texttt{union}, \texttt{intersection}, and \texttt{difference}. Each of these set-set operations is equivalent to batched operations: the data structure should process a set of operations \texttt{insert}, \texttt{delete} and \texttt{contains}. It is obvious that we want these ``large'' operations to be parallelized. Typically, these sets are implemented with the trees of logarithmic height, such as 2-3 tree, Treap, AVL tree, Red-Black tree, etc. Until now, little attention was devoted to data structures that work better but under several restrictions on the data. In this work, we parallelize Interpolation Search Tree which serves each request from a \emph{smooth} distribution in doubly-logarithmic time. Our data structure of size $n$ performs a batch of $m$ operations in $O(m \log\log n)$ work and poly-log span.
\end{abstract}

\maketitle

\vspace{-0.2cm}
\section{Preliminaries}

\paragraph{Model.} We use the work-span model to calculate the complexity of our algorithm.

\vspace{-0.1cm}
\paragraph{Related work.} There are several works that presented parallel batched search trees. Paul, Vishkin and Wagener~\citep{paul1983parallel} studied 2-3 trees.
Park and Park~\citep{park2001parallel} showed similar results for red-black trees.
Blelloch and Reid-Miller~\citep{blelloch1998fast} presented a work-efficient parallel batched treap.
Akhremtsev and Sanders~\citep{akhremtsev2016fast} looked on work-efficient $(a,b)$-trees.
And, finally, Blelloch et al.~\citep{blelloch2016just} worked on the parallel work-efficient batched generic binary search trees with the only provided \emph{Join} function.
However, none of these works consider the trees that could have lower than logarithmic height under a wide range of distributions.
In this work, we consider the parallel batched version of Interpolation Search Tree (later called IST) introduced in~\citep{mehlhorn1993dynamic}.
The main difference with the previously researched data structures is that insertions, deletions and searches with smoothly distributed arguments take $O(\log \log n)$ time, where $n$ is the size of the tree.

\vspace{-0.1cm}
\paragraph{Standard functions.} We need to use several standard \emph{parallelizable} functions. \emph{Span} function calculates the prefix sums of the array of size $n$ in $O(n)$ work and $O(\log n)$ span. \emph{Filter} function filters out all the elements that do not satisfy the condition $C$. If the size of the array is $n$, it takes $O(n \cdot \mathrm{cost}(C))$ work and $O(\log n \cdot \mathrm{cost}(C))$ span. Please note that in our data structure condition functions are pretty simple and they work in $O(1)$, thus, the complexity of filter becomes $O(n)$ work and $O(\log n)$ span.
Also, we need \emph{rank} function that given two sorted arrays $a$ and $b$ of sizes $n$ and $m$, correspondingly, finds the position $k$ of each element $a[i]$ in $b$ such that $b[k - 1] \leq a[i] \leq b[k]$. It works in $O(n + m)$ work and $O(\log^2 (n + m))$ span.
Moreover, we need \emph{merge} algorithm that given two sorted arrays $a$ and $b$ of size $n$ and $m$, correspondingly, merges to arrays together with $O(n + m)$ work and $O(\log^2 (n + m))$ span.  \footnote{We can just use \emph{rank} to achieve such merge complexity.} Finally, we need a \emph{parallel-for} loop on $n$ iterations that executes each loop step in parallel. It is implemented using a binary-splitting technique and introduces additional $O(\log n)$ span and $O(n)$ work.

\vspace{-0.1cm}
\paragraph{Result.} Our data structure, parallel batched IST, performs a batch of $M$ requests in expected $O(M \log \log (n + M))$ work and $\mathrm{polylog}(n + m)$ span. It outperforms parallel batched Treap presented in~\citep{blelloch2016just} on our experiments.

\vspace{-0.2cm}
\section{IST definition}
\begin{definition}
Let $a$ and $b$ are reals. An Interpolation Search Tree (IST) with boundaries $a$ and $b$ for a set of $n$ keys $\{x_1, \ldots, x_n\} \subseteq [a, b]$ consists of:
\begin{enumerate}
    \item An array REP of representatives $x_{i_1}, \ldots, x_{i_k}, i_1 < i_2 < \ldots < i_k$ that is $\text{REP}[j] = x_{i_j}$. Furthermore, $k$ satisfies $\sqrt{n} / 2 \leq k \leq 2\cdot\sqrt{n}$.
    \item Interpolation search trees $T_1, \ldots, T_{k+1}$ for subarrays of keys $S_1, \ldots, S_{k + 1}$ where $S_j = \{x_{i_{j-1}+1}, \ldots x_{i_j - 1}\}$ for $2 \leq j \leq k$, $S_1 = \{x_1, \ldots, x_{i_1-1}\}$, and $S_{k + 1} = \{x_{i_k+1}, \ldots, x_{n}\}$. Furthemore, tree $T_j$, $2 \leq j \leq k$ has boundaries $x_{i_{j-1}}$ and $x_{i_j}$, $T_1$ has boundaries $a$ and $x_1$, and $T_{k + 1}$ has boundaries $x_k$ and $b$.
    \item An array $\text{ID}[1, \ldots, m]$, where $m$ is some integer, with $\text{ID}[i] = j$ iff $\text{REP}[j] < a + i(b - a) / m \leq \text{REP}[j + 1]$.
\end{enumerate}
\end{definition}

Array REP contains a sample of the keys and array ID helps to find a value in the array REP, i.e., it gives the first approximation that is then corrected to the actual position, using some search technique, for example, binary search. We require \emph{ideal} IST to have the samples in REP to be equally spaced. We rebuild subtrees in insertion and deletion algorithms at \emph{appropriate} time intervals, i.e. when enough \texttt{insert} or \texttt{delete} operations have been executed on that subtree.

\begin{definition}
An IST with parameter $\alpha$, $\frac{1}{2} \leq \alpha < 1$, for set $S$ of size $n$ is \emph{ideal} if $i_j = j [\sqrt{n}]$ for all $j \geq 1$, if $m = [n^\alpha]$, and if the subtrees are again ideal ISTs.
\end{definition}

In the ideal IST the first level node contains $\sqrt{n}$ keys, all nodes on the second level contains $\sqrt[4]{n}$ keys, and so on.

\begin{theorem}
\label{thm:ideal}
Let $\frac{1}{2} \leq \alpha < 1$. Then an ideal IST for an ordered set $S$ of size $n$ can be built in $O(n)$ time and require $O(n)$ space. It has depth $O(\log \log n)$.
\end{theorem}

Additionally, at each subtree $T$ we maintain the number of operations from the last rebuild $C(T)$, the size of the subtree just after the last rebuild $S_0(T)$, and the current size of the subtree $S(T)$.

Thus, an insertion and a deletion work as follows. They traverse the IST recursively, visiting in each node the desired child, also they maintain the counters $C(T)$ and $S(T)$ until it finds a node with $C(T) \geq S_0(T) / 4$. At that node it builds a new \emph{ideal} IST out of $T$ including the element, being inserted. Deletions are performed by marking the element to be deleted. The element is physically removed only during the rebuild operation.

\begin{theorem}[\citep{mehlhorn1993dynamic}]
Let $\mu$ be a smooth density of finite support $[a, b]$ with parameter $\alpha$, $\frac{1}{2} \leq \alpha < 1$, and let $T$ be a $\mu$-random IST of size $n$. Then, the expected cost of processing a $\mu$-random search is $O(\log \log n)$ and $O(\log^2 n)$ worst time. The expected amortized cost of processing a $\mu$-random insertion and deletion is $O(\log \log n)$ and $O(\log n)$ worst case amortized time.
\end{theorem}

\section{Parallel Construction of Ideal IST}
We start the description of our algorithm with how to build an ideal IST from the sorted array $a$ of keys with length $n$.

For that we choose each $[\sqrt{n}]$-th element ($[\sqrt{n}], 2[\sqrt{n}], \ldots$) and create an array REP out of them. Then, recursively build the ideal ISTs from subarrays $a[1, \ldots, [\sqrt{n}]-1]$, $a[[\sqrt{n}] + 1, 2[\sqrt{n}]-1]$, and so on. To build them we used a parallel-for loop with $O(\sqrt{n})$ iterations giving $O(\sqrt{n})$ work and $O(\log n)$ span. Finally, we have to find array ID of length $m = n^{\alpha}$, i.e., for each $k$ find the position of $a + k \cdot (b - a) / m$ in array REP. For that we can use the standard \emph{rank} algorithm with $O(n + m) = O(m)$ work and $O(\log^2 (n + m)) = O(\log^2 n)$ span.

Our rebuild algorithm runs recursively on $O(\log \log n)$ levels, since the height of an ideal tree is $O(\log \log n)$. Thus, in total the work is $O(n)$ by Theorem~\ref{thm:ideal} and the span is $O(\log \log n \cdot \log^2 n)$.

\section{Flatten IST into Array}

The other thing that we need for our data structure is to flatten the tree: get all the unmarked keys of the tree into the sorted array. For that we use the counters $S$ for the size of subtrees. We create an array with the length of REP array and calculate prefix sums of the sizes of the subtrees using \emph{scan} function while \emph{filtering} out the marked keys in REP array. Then, we recursively go into subtrees knowing the position where they should be located in the array. The height of IST is $O(\log n)$ at worst. Thus, this algorithm works in $O(n)$ work and $O(\log^2 n)$ span.

\section{Batched operations}
Now, we are given a batch of $M$ operations that we want to apply to a $\mu$-random IST $T$. At first, we compare how many operations was done to $T$ since the last rebuild $C(T)$ with the size of $T$, $S_0(T)$, after the last rebuild divided by $4$, i.e., $S_0(T) / 4$. If the number of operations is smaller, then we simply find for each operation in which subtree it should go using standard search algorithm in REP array of IST, mark the necessary elements to be inserted or deleted, and then continue recursively with each subtree. Also, we update the size of the subtree. Otherwise, if the number of operations since the last rebuild is bigger, then we transform the tree into a sorted array using the previously described algorithm, merge it with the keys of operations, apply the operations using \emph{parallel-for} loop, and, finally, build the tree from a sorted array using the previously described algorithm. Since the height of IST is at most $O(\log n)$, the work is expected to be $O(M \log \log (n + M))$ and the span is $O(\mathrm{polylog} (n + M))$.

\vspace{-0.2cm}
\section{Experiments}
We implement the data structure on C++ and compile it with OpenCilk 1.0 compiler~\citep{opencilk} using \texttt{-O2} flag. We wrote the code with the help of the pctl library presented in~\citep{acar2019provably}. It contains the scalable implementations of standard functions and provides an automatic solution to the granularity problem.
Also, to resolve the issue with the parallel memory allocation we used \texttt{tcmalloc}~\citep{tcmalloc}.

We run our code on an Intel Xeon Gold 6230 machine with 16 cores.

We compared our parallel batched IST with the parallel batched Treap explained in~\citep{blelloch2016just} and the sequential std::set from C++ standard library.
For the experiments, we fill the data structures with approximately $10^7$ of keys uniformly taken from $[1, 2 \cdot 10^7]$ (each key is taken with the probability $1/2$) and then we apply the batch of $10^6$ insertions with the keys taken uniformly from $[1, 2 \cdot 10^7]$. We run data structures on $1$ and $16$ processes and calculate their speedup. The experimental results are presented in the following table. Each value was averaged over $10$ runs.

\begin{center}
\begin{tabular}{c|c|c|c}
& 1 proc, s & 16 proc, s & speedup \\\hline
IST & 5.1s & 0.36s & 14.1 \\
Treap & 7.5s & 0.5s & 14.9 \\
std::set & 3.6s & --- & --- \\
\end{tabular}
\end{center}

As one can see, our data structure outperforms std::set on 16 processes and outperforms parallel batched Treap on both settings, however, it has a little bit worse speedup.

\vspace{-0.2cm}
\section{Conclusion} In this short paper, we presented a new parallel batched data structure based on the sequential IST~\citep{mehlhorn1993dynamic}. It appears to perform better than parallel batched Treap presented in~\citep{blelloch2016just}. In the future, we want to run more experiments on the trees of different sizes with different types of operations and on different machines with more cores.

\bibliography{references}

\end{document}